\documentclass[english]{article}
\usepackage{amsmath}
\usepackage[T1]{fontenc}
\usepackage{babel}
\usepackage{amsmath}
\usepackage{graphicx}
\usepackage{fancyhdr}
\usepackage{cite}
\usepackage{url}

\newcommand{\pset}{\vartheta}

\newcommand{\likelihood}{p(\vec{d}(t)|\vec{\pset},H)}
\newcommand{\prior}{p(\vec{\pset}|H)}
\newcommand{\evidence}{p(\vec{d}(t)|H)}
\newcommand{\posterior}{p(\vec{\pset}|\vec{d}(t),H)}

\graphicspath{{./}{figures/}}

\date{}

\begin{document}

\title{Posterior samples of the parameters of binary black holes from Advanced LIGO, Virgo's second observing run}



\author{Soumi De\textsuperscript{1{*}}
Christopher M. Biwer\textsuperscript{2}
Collin D. Capano\textsuperscript{3,4} \\
Alexander H. Nitz\textsuperscript{3,4}
Duncan A. Brown\textsuperscript{1}}

\maketitle
\thispagestyle{fancy}

\begin{center}
1. Department of Physics,
Syracuse University, Syracuse NY 13244, USA \\
\vspace{0.1cm}
2. Computer, Computational, and Statistical Sciences Division, Los Alamos National Laboratory, Los Alamos, NM 87545, USA \\
\vspace{0.1cm}
3. Albert-Einstein-Institut, 
Max-Planck-Institut f\"ur Gravitationsphysik, 
D-30167, Hannover, Germany \\
\vspace{0.1cm}
4. Leibniz Universit{\"a}t Hannover, D-30167, Hannover, Germany \\
\vspace{0.1cm}
{*}corresponding author:
Soumi De (sde101@syr.edu)
\end{center}

\begin{abstract}
This paper presents a parameter estimation analysis of the seven binary black hole mergers---GW170104, GW170608, GW170729, GW170809, GW170814, GW170818, and GW170823---detected during the second observing run of the Advanced LIGO and Virgo observatories using the gravitational-wave open data. We describe the methodology for parameter estimation of compact binaries using gravitational-wave data, and we present the posterior distributions of the inferred astrophysical parameters. We release our samples of the posterior probability density function with tutorials on using and replicating our results presented in this paper.
\vspace{1cm}
\end{abstract}

\section*{Background \& Summary} \label{sec:intro}

During the second Advanced LIGO--Virgo observing run (O2), three binary black hole mergers were observed by the Advanced LIGO detectors~\cite{TheLIGOScientific:2014jea} on January 4, 2017---GW170104~\cite{Abbott:2017vtc}, June 8, 2017---GW170608~\cite{Abbott:2017gyy}, and August 23, 2017---GW170823~\cite{LIGOScientific:2018mvr} and four binary black hole mergers observed by the Advanced LIGO detectors and the Advanced Virgo detector\cite{TheVirgo:2014hva} on July 29, 2017---GW170729~\cite{LIGOScientific:2018mvr}, August 9, 2017---GW170809~\cite{LIGOScientific:2018mvr}, August 14, 2017---GW170814~\cite{Abbott:2017oio} and August 18, 2017---GW170818~\cite{LIGOScientific:2018mvr}. Including the binary black hole mergers observed in Advanced LIGO's first observing run~\cite{TheLIGOScientific:2016pea,Nitz:2018imz} (O1), to date, there have been ten binary black hole mergers reported to have been detected by the Advanced LIGO--Virgo observatories~\cite{TheLIGOScientific:2016pea,Nitz:2018imz,Abbott:2017vtc,Abbott:2017gyy,Abbott:2017oio,LIGOScientific:2018mvr}. The properties of these observed binary black hole sources (eg. masses and spins) are of interest to the astrophysics community to understand the formation, evolution, and populations of black holes. These properties are estimated using Bayesian inference~\cite{Bayes:1763,Jaynes:2003jaq} which allow us to sample the posterior probability density function---the probability of the modeled parameter values given a model and set of detectors' data. We perform a Bayesian inference analysis~\cite{Christensen:2001cr,Biwer:2018osg} using the available gravitational-wave data~\cite{Vallisneri:2014vxa} for GW170104, GW170608, GW170729, GW170809, GW170814, GW170818, and GW170823---the seven binary black holes reported from O2, and we present their posterior probability density functions in this paper. In particular, we present estimates for the masses, spins, distances, inclination angle, and sky locations of the binaries.

\section*{Methods}\label{sec:pe_analysis}

\subsection*{Bayesian inference}

We perform a Bayesian parameter estimation analysis~\cite{Christensen:2001cr} to measure the source properties of the seven binary black-mergers from Advanced LIGO--Virgo's second observing run, using the gravitational-wave data available at the Gravitational-Wave Open Science Center~\cite{Vallisneri:2014vxa}. We use the data available from the Advanced LIGO detectors for GW170104, GW170608, GW170823. For GW170729, GW170809, GW170814, and GW170818, we use the available Advanced LIGO and the Advanced Virgo data. The parameter estimation analysis was executed using the PyCBC Inference software \cite{Biwer:2018osg,alex_nitz_2018_1410598} and the parallel-tempered emcee sampler \cite{emcee,vousden:2016,mcmc}, which employs ensemble Markov chain Monte Carlo (MCMC) techniques~\cite{metropolis1953,geman:1984,Gelman95bayesiandata,Christensen:2001cr,Christensen:2004bc,Rover:2006ni,Rover:2006bb,TheLIGOScientific:2016pea,Abbott:2017vtc,Abbott:2017gyy,Abbott:2017oio} to sample the posterior probability density function $\posterior$. We calculate the posterior probability density function, $\posterior$, for the set of parameters $\vec{\pset}$ for the gravitational-waveform model, $H$, given the gravitational-wave data from the detectors $\vec{d}(t)$~\cite{Vallisneri:2014vxa}
\begin{equation}
\label{eqn:bayes} \posterior =
\frac{\likelihood \prior}{\evidence} ,
\end{equation}
where $\prior$ is the prior---the assumed knowledge of the distributions for the parameters $\vec{\pset}$ describing the signal, before considering the data. $\likelihood$ is the likelihood---the probability of obtaining the data $\vec{d}(t)$ given the model $H$ with parameters 
$\vec{\pset}$. The likelihood in a network of $N$ detectors is computed as~\cite{Biwer:2018osg,Wainstein:1962,Rover:2006bb} 
\begin{equation}\label{eqn:log_likelihood}
\likelihood = \\
\exp \left[ -\frac{1}{2} \sum_{i = 1}^{N} \left<\tilde{d}_{i}(f) - \tilde{s}_{i}(f, \vec{\pset})| \tilde{d}_{i}(f) - \tilde{s}_{i}(f, \vec{\pset})\right> \right]
\end{equation}
considering the noise in each detector to be stationary, Gaussian, and uncorrelated with the noise in the other detectors in the network. $\tilde{d}_{i}(f)$, $\tilde{n}_{i}(f)$, and $\tilde{s}_{i}(f, \vec{\pset})$ are the
frequency-domain representations of the data, noise, and the model waveforms respectively. The inner product $\langle\tilde{a} | \tilde{b}\rangle$ is defined as 
\begin{equation}
\left<\tilde{a}_i(f) | \tilde{b}_i(f)\right> = 4\Re \int_{0}^{\infty} \frac{\tilde{a}_i(f) \tilde{b}_i(f)}{S^{(i)}_n(f)} \mathrm{d}f \,,
\end{equation}
where $S^{(i)}_n(f)$ is the power spectral density (PSD) of the $i$-th detector's noise.

For computing the likelihood, we analyze the  gravitational-wave dataset $\vec d(t)$ from the Hanford and Livingston detectors, between GPS times (1167559926, 1167559942) for GW170104, (1180922444, 1180922500) for GW170608, and (1187529246, 1187529262) for GW170823. We analyze $\vec d(t)$ from the Hanford, Livingston, and Virgo detectors between GPS times (1185389797, 1185389813) for GW170729, (1186302509, 1186302525) for GW170809, (1186741851, 1186741867) for GW170814, and (1187058317, 1187058333) for GW170818. Based on the estimates of the masses indicating the length of the signals from the search pipeline~\cite{alex_nitz_2018_1410598,Usman:2015kfa,Canton:2014ena,Nitz:2017svb,TheLIGOScientific:2016qqj} and from the results of the parameter estimation analyses reported in Refs.~\cite{Abbott:2017gyy,Abbott:2017vtc,Abbott:2017oio,LIGOScientific:2018mvr}, GW170608 was found to have properties of a lower mass source and hence have larger number of cycles as compared to the other events. Therefore we extend the priors for GW170608 to much lower component masses than for the other two events, which is described below.
This requires more data for the analysis of GW170608 such that the segment of the data being analyzed can encompass the longest duration (ie. smallest mass) template waveform drawn from the prior used for GW170608.

The dataset is decimated to a sample rate of 2048~Hz. The PSD used in the likelihood is constructed using the median PSD estimation method described in Ref.~\cite{Allen:2005fk} with 8~s Hann-windowed segments (overlapped by 4~s) taken from GPS times (1167559424, 1167560448) for GW170104, (1180921982, 1180923006) for GW170608, (1185388936, 1185389960) for GW170729, (1186302007, 1186303031) for GW170809, (1186741349, 1186742373) for GW170814, (1187057815, 1187058839) for GW170818, and (1187528744, 1187529768) for GW170823. Prior to performing a Fourier transform of the data for PSD estimation, we remove the signal from the data used for PSD estimation by applying a gating window of width of the order of the signal length. This removes any bias introduced in the noise due to the presence of the signal. The PSD estimate is truncated to 4~s in the time-domain using the method described in Ref.~\cite{Allen:2005fk}.
For all seven events except GW170608, the likelihood is computed between a low-frequency cutoff of 20~Hz and the Nyquist frequency of 1024~Hz for all the detectors in the network. For GW170608, we use the same procedure in Ref.~\cite{Abbott:2017gyy} and compute the likelihood using a low-frequency cutoff of 20~Hz and the Nyquist frequency of 1024~Hz for the Livingston detector, and using frequencies between 30~Hz and 1024~Hz for the Hanford detector. During the observation of GW170608, the Hanford detector was undergoing a routine instrumental procedure to minimize angular noise coupling to the strain measurement. This introduced excess noise in the strain data from the Hanford detector at frequencies around $\sim$19-23~Hz, but the strain data was shown to be stable above 30~Hz in Ref.~\cite{Abbott:2017gyy}.  

The template waveforms $\tilde{s}_{i}(f, \vec{\pset})$ used in the likelihood are generated using the IMRPhenomPv2~\cite{Schmidt:2014iyl,Hannam:2013oca} waveform model implemented in the LIGO Algorithm Library (LAL)~\cite{lallib}. The parameters $\vec{\pset}$ measured in the ensemble MCMC for these seven events are: right ascension $\alpha$, declination $\delta$, polarization $\psi$, component masses in the detector frame $m_1^{\mathrm{det}}$ and $m_2^{\mathrm{det}}$, luminosity distance $d_L$, inclination angle $\iota$, coalescence time $t_c$, magnitudes for the spin vector $a_1$ and $a_2$, azimuthal angles for the spin vectors $\theta_1^\mathrm{a}$ and $\theta_2^\mathrm{a}$, polar angles for the spin vectors $\theta_1^\mathrm{p}$ and $\theta_2^\mathrm{p}$. We analytically marginalize over the fiducial phase $\phi$. For efficient sampling of the parameter space and faster convergence of the Markov chains, we apply a transformation from the mass parameters that define the prior ($m_1^{\mathrm{det}}$, $m_2^{\mathrm{det}}$) to chirp mass and mass ratio $(\mathcal{M}^{\mathrm{det}}, q)$ coordinates. The chirp mass is defined as $\mathcal{M} = (m_1 m_2)^{3/5}/(m_1 + m_2)^{1/5}$. While sampling, we allow the mass ratio $q$ to be both greater and less than 1.

For GW170104, we assume uniform priors for detector-frame component masses $m_{1,2}^{\mathrm{det}} \in$ [5.5, 160) M$_\odot$. When generating the waveform in the MCMC, the masses are transformed to the detector-frame chirp mass $\mathcal{M}^\mathrm{det}$ and $q$ with a restriction $12.3 < \mathcal{M}^\mathrm{det}/M_\odot < 45.0$, and $1 < q < 8$ where $q = \max\{ m_1^{\mathrm{det}}, m_2^{\mathrm{det}} \} / \min\{ m_1^{\mathrm{det}}, m_2^{\mathrm{det}} \}$. We assume uniform prior distributions $m_{1,2}^{\mathrm{det}} \in$ [3, 50) M$_\odot$ for GW170608, $m_{1,2}^{\mathrm{det}} \in$ [10, 90) M$_\odot$ for GW170729, $m_{1,2}^{\mathrm{det}} \in$ [10, 80) M$_\odot$ for GW170814, and $m_{1,2}^{\mathrm{det}} \in$ [5, 80) M$_\odot$ for GW170809, GW170818, and GW170823. For the luminosity distance, we assume a uniform in volume distribution such that $p(d_L | H)~\propto d_L^2$, with $d_L \in$ [100, 2500) Mpc for GW170104, $d_L \in$ [10, 1500) Mpc for GW170608, $d_L \in$ [10, 5000) Mpc for GW170729, $d_L \in$ [10, 2500) Mpc for GW170809, $d_L \in$ [10, 1500) Mpc for GW170814, $d_L \in$ [10, 3000) Mpc for GW170818, and $d_L \in$ [10, 5000) Mpc for GW170823. The priors for the remaining parameters are the same for all the events. For spin magnitudes, we use uniform priors $a_{1,2} \in$ [0.0, 0.99).
We use a uniform solid angle prior for the spin angles, assuming a uniform distribution for the spin azimuthal angles $\theta_{1,2}^\mathrm{a} \in [0, 2\pi)$ and a sine-angle distribution for the spin polar angles $\theta_{1,2}^\mathrm{p}$.
We use uniform priors for the arrival time $t_c \in [t_s - 0.1~s, t_s + 0.1~s)$ where $t_s$ is the trigger time of the event being analyzed, reported in ~\cite{Abbott:2017vtc,Abbott:2017gyy,Abbott:2017oio}.
For the sky location parameters, we use a uniform distribution prior for $\alpha \in [0, 2\pi)$ and a cosine-angle distribution prior for $\delta$. We use a uniform prior for the polarization angle $\psi \in [0, 2 \pi)$ and a sine-angle distribution for the inclination angle $\iota$ prior. The mass and spin priors for GW170104 are the same as those mentioned for the final analysis using the ``effective precession'' model in Ref.~\cite{Abbott:2017vtc}.

The parameter estimation analyses of the events produce samples of the posterior probability density function in the form of Markov chains. Successive states of these chains are not independent, as Markov processes depend on the previous state~\cite{Christensen:2004jm}. Independent samples are obtained from the full Markov chains by ``thinning'' or drawing samples from chains of the coldest temperature, with an interval of the autocorrelation length~\cite{Biwer:2018osg,Christensen:2004jm}. These independent samples are used to calculate estimates for the model parameters from the analysis.

\subsection*{Posterior probability density functions}\label{sec:results}

Independent samples from the ensemble MCMC chains from the analyses of all the seven events are available for download at the data release repository for this work~\cite{data_release}. We encourage use of these data in derivative works. The repository also contains IPython notebooks \cite{PER-GRA:2007} demonstrating how to read the data from the files and manipulate them, and provide examples of reconstructing the figures presented in this paper.

Samples of the varied parameters in the MCMC can be combined to obtain posteriors for other derivable parameters. We map the values for the detector-frame masses ($m_{1}^{\mathrm{det}}$, $m_{2}^{\mathrm{det}}$) and the luminosity distance $d_L$ from the runs to source-frame masses ($m_{1}^{\mathrm{src}}$, $m_{2}^{\mathrm{src}}$) using the standard $\Lambda$-CDM cosmology~\cite{Schutz:1986gp,Finn:1992xs}. While visualizing and quoting the detector-frame and source-frame masses, we use $q = m_1^\mathrm{det} / m_2^\mathrm{det} =  m_1^\mathrm{src} / m_2^\mathrm{src}$ where $m_1^\mathrm{det}$ and $m_1^\mathrm{src}$ refer to the more massive black hole, and $m_2^\mathrm{det}$ and $m_2^\mathrm{src}$ refer to the less massive black hole in the binary; ie. we present our results with $q \geq 1$. We also map the component masses to parameters such as the chirp mass $\mathcal{M}$ and the mass ratio $q$, and map the component masses and spins to the effective inspiral spin parameter $\chi_{\mathrm{eff}}$ and the effective precession spin parameter $\chi_p$~\cite{Schmidt:2014iyl,Hannam:2013oca}. Our measurements show that all the events are in agreement with being binary black hole sources.

\begin{figure}[ht]
  \includegraphics[width=\textwidth]{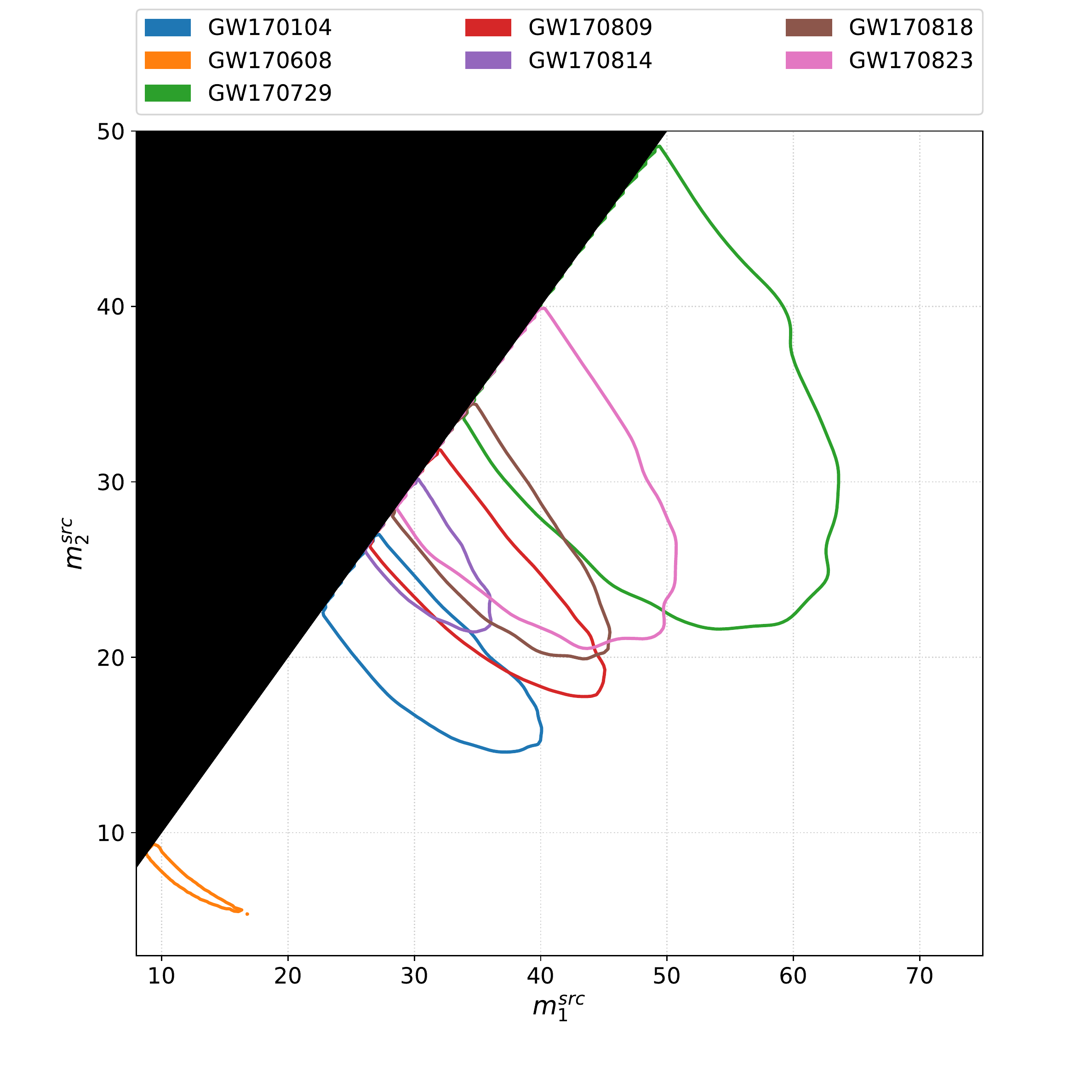}
  \caption{Posterior probabilities of the source frame primary mass $m_1^{\mathrm{src}}$ and secondary mass $m_2^{\mathrm{src}}$ from the PyCBC Inference analyses of the seven gravitational-wave signals from binary black-hole mergers in Advanced LIGO-Virgo's second observing run (O2). Plotted are the 90\% credible contours in the 2D plane. 
The measurements suggests that GW170729 has the highest masses and GW170608 has the lowest masses among all black hole binaries observed in O1 and O2. Parameter estimates of the O1 binary black holes were presented in Refs.~\cite{Biwer:2018osg,TheLIGOScientific:2016pea}.\label{fig:m1m2_plots}} 
\end{figure}

\begin{figure}[ht]
  \includegraphics[width=\textwidth]{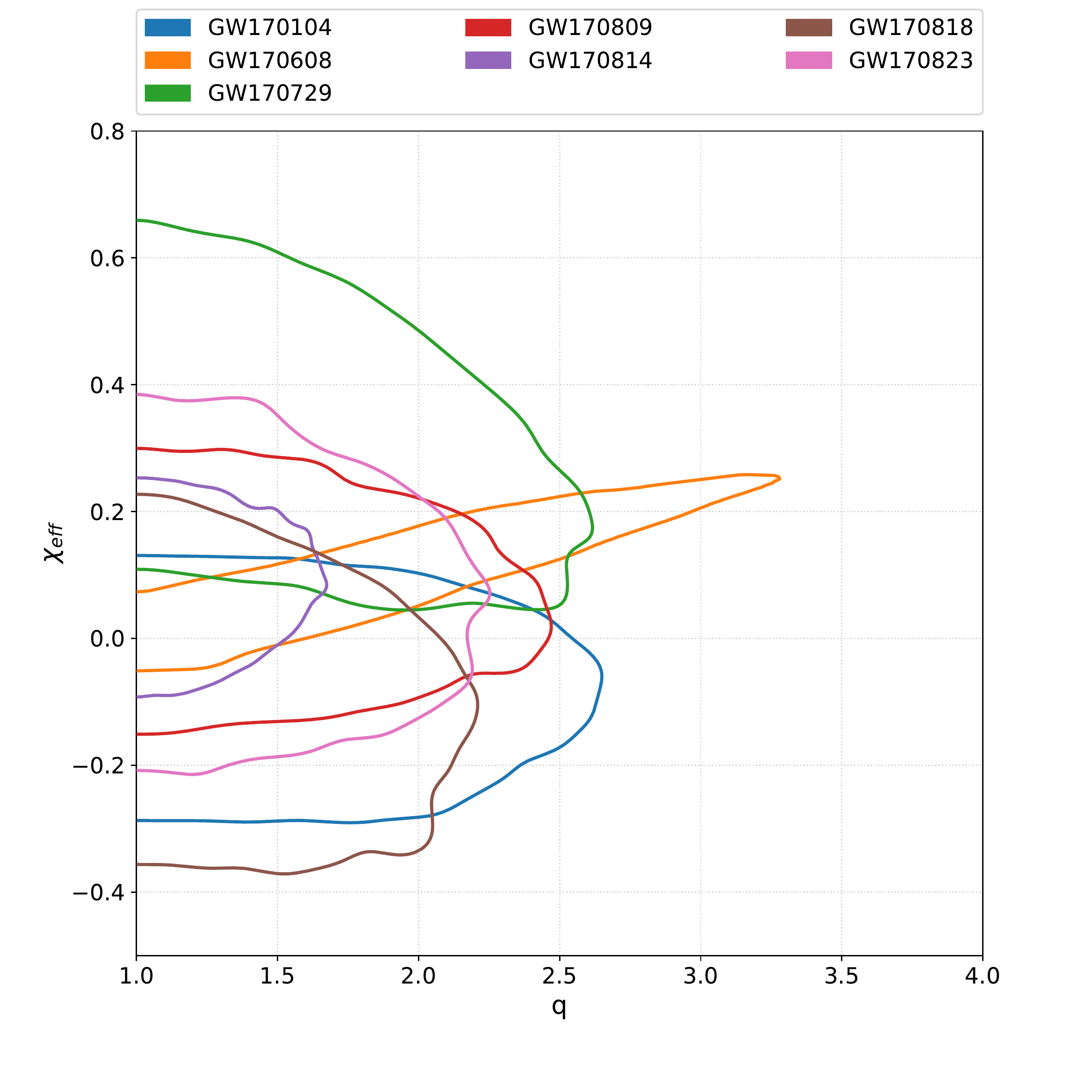}
  \caption{Posterior probabilities of the asymmetric mass ratio $q$ and the effective inspiral spin $\chi_{\mathrm{eff}}$ from the PyCBC Inference analyses of the seven gravitational-wave signals from binary black-hole mergers in Advanced LIGO-Virgo's second observing run. Plotted are the 90\% credible contours in the 2D plane. All the events have low $\chi_{\mathrm{eff}}$ values. GW170814 has lesser support for asymmetric mass ratios than the other events.
  \label{fig:qchieff_plots}
}
\end{figure}

\begin{figure}[ht]
  \includegraphics[width=\textwidth]{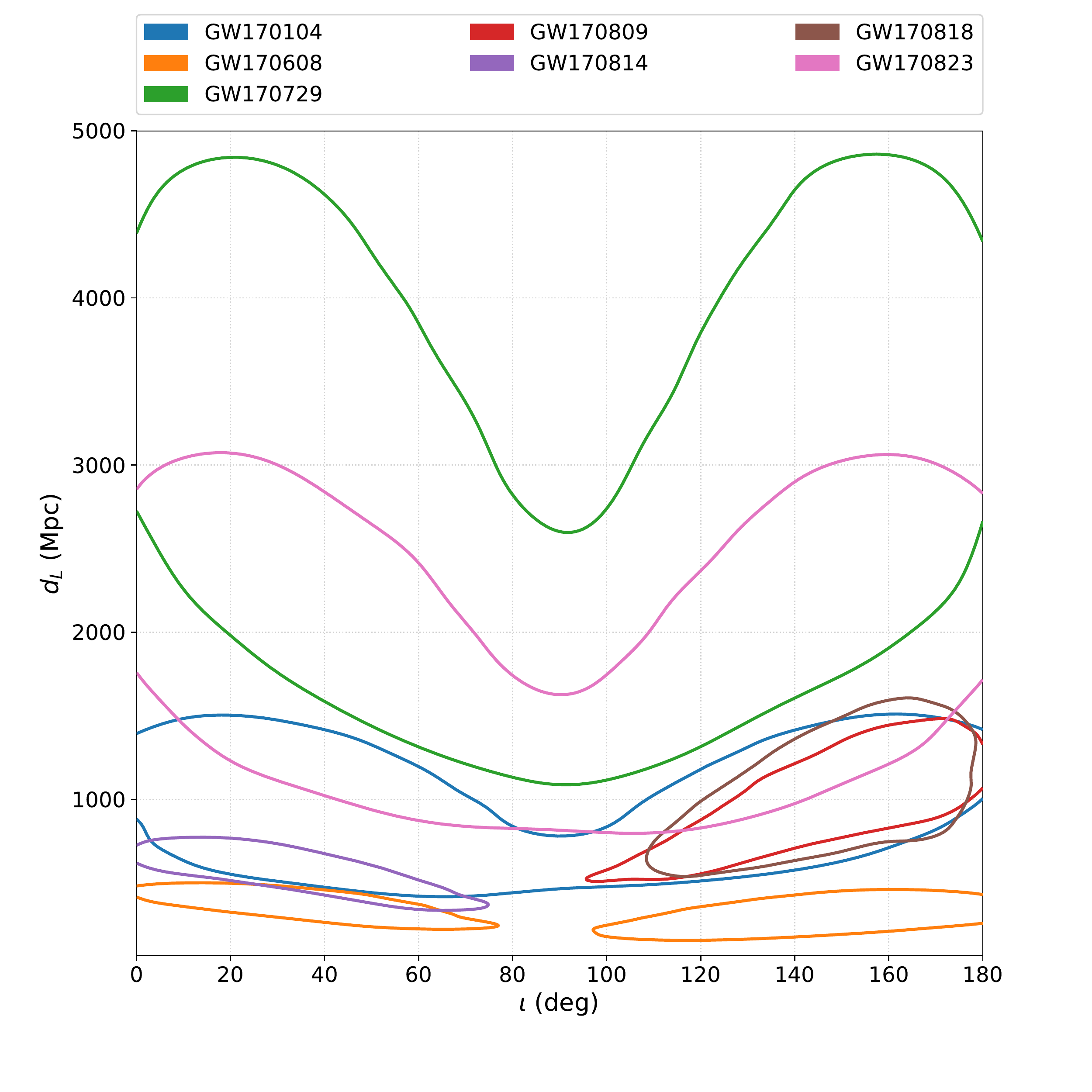}
  \caption{Posterior probabilities of the luminosity distance $d_L$ and the inclination angle $\iota$ from the PyCBC Inference analyses of the seven gravitational-wave signals from binary black-hole mergers in Advanced LIGO-Virgo's second observing run. Plotted are the 90\% credible contours in the 2D plane. GW170104, GW170729, and GW170823 have support for face-on ($\iota = 0$), face-off ($\iota = 180$) and edge-on ($\iota = 90$). For GW170608, there is a stronger preference for the system being face-on ($\iota = 0$) and face-off ($\iota = 180$). For GW170814, there is a stronger preference for the system being face-on ($\iota = 0$). For GW170809 and GW170818 there is a stronger preference for  face-off ($\iota = 180$). GW170608 is observed at the closest luminosity distance and GW170729 the farthest. \label{fig:dliota_plots}
}
\end{figure}

In order to obtain an estimate for a particular parameter, the other parameters that were varied in the ensemble MCMC can be marginalized over in the posterior probability density function. Recorded in Table~\ref{tab:results}, is a summary of the median and 90\% credible interval values of the main parameters of interests obtained from the analyses of all seven O2 binary black hole events.
The marginalized distributions for $m_1^{\mathrm{src}} - m_2^{\mathrm{src}}$, $q - \chi_{\mathrm{eff}}$, and $d_L - \iota$ for the seven events are shown in  Figs.~\ref{fig:m1m2_plots}, \ref{fig:qchieff_plots}, and \ref{fig:dliota_plots} respectively.
The two-dimensional plots in these figures show 90\% credible regions for the respective parameters.

Our results show that GW170729 is the largest mass binary black hole signal and GW170608 is the smallest mass binary black hole signal from the detections during O1 and O2. Parameter estimates of the binary black holes observed during O1 were presented in Refs.~\cite{Biwer:2018osg, TheLIGOScientific:2016pea}.
GW170814 seems to have lesser support for asymmetric mass ratios than the other events. All the events have low effective spin values. GW170814 has more support for face-on systems, whereas GW170809 and GW170818 has a preference for face-off systems. For GW170608, there is preference for both face-on ($\iota = 0$) and face-off ($\iota = 180$). GW170104, GW170729, and GW170823 has support for face-on ($\iota = 0$), face-off ($\iota = 180$) and edge-on ($\iota = 90$). Face-on systems are those for which the inclination angle $\iota = 0$; ie. the line of sight is parallel to the binary's orbital angular momentum. Face-off systems are those for which $\iota = \pi$ (the line of sight is anti-parallel to the binary's orbital angular momentum). We also computed $\chi_{p}$ for each of the events and found no significant measurements of precession. GW170608 seems to be observed at the closest luminosity distance and GW170729 the farthest among the O2 binary black holes.

\begin{figure}[ht]
  \includegraphics[width=\textwidth]{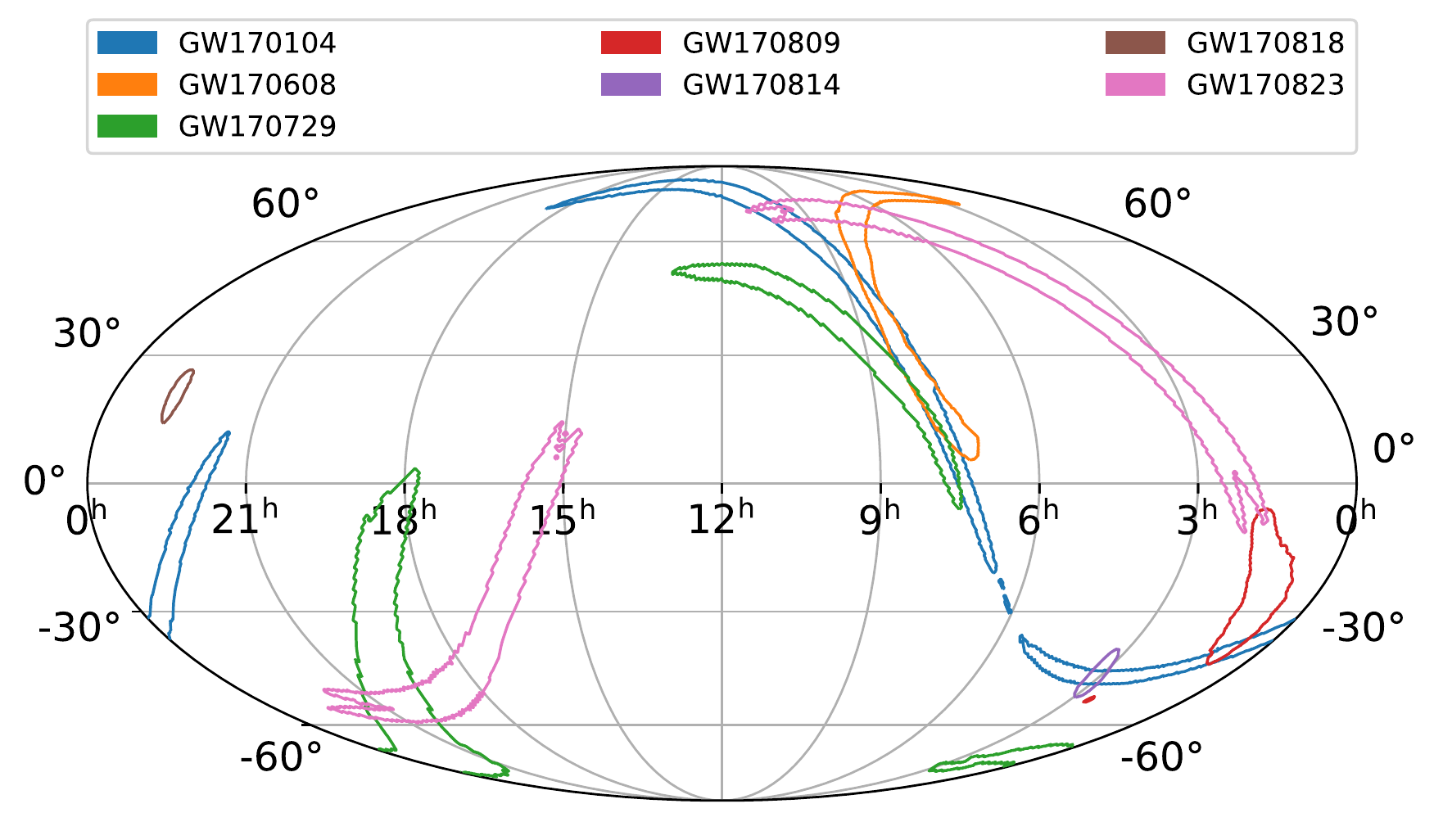}
  \caption{Posterior probabilities for the sky location parameters---right ascension and declination from the PyCBC Inference analyses of the seven gravitational-wave signals from binary black-hole mergers in Advanced LIGO-Virgo's second observing run. Plotted are the 90\% credible contours in Mollweide projection and celestial coordinates; the right ascension is expressed in hours and the declination in degrees. GW170818 and GW170814 have substantially small sky localization areas, being detected by the H1L1V1 three-detector network, with considerable SNR in all the detectors. The GW170729 and GW170809 analyses used data from the three-detector network. However, the sky localization area is broad due to low Virgo SNR. Between GW170729 and GW170809, the latter has a higher network SNR leading to a smaller sky localization area. GW170104, GW170608, GW170823 have poor sky localization, as they were detected by the H1L1 two detector network; GW170823 has the lowest network SNR and broadest sky localization area, and GW170608 has the highest network SNR causing the smaller sky localization area. \label{fig:skymap_plots}
 }
\end{figure}

Figs.~\ref{fig:skymap_plots} shows the 90\% credible regions for the sky location posterior distributions of all the seven binary black-hole events in a Mollweide projection and celestial coordinates. GW170818 and GW170814 have substantially small sky localization areas as they were detected by the H1L1V1 three-detector network, with a significant signal-to-noise ratio (SNR) contribution from all the detectors. The GW170729 and GW170809 parameter estimation analyses use data from all three detectors in the network. However, the SNR in Virgo is not significant, causing the sky localization area to be broader than in the cases of GW170814 and GW170818.  The sky localization area of GW170809 is smaller as compared to GW170729, as the former has a higher network SNR than the latter; the sky localization area varies inversely as the square of the SNR. The events observed by the H1L1 two-detector network---GW170104, GW170608, GW170823 have poor sky localization, with GW170823 having the lowest network SNR and broadest sky localization area, and GW170608 having the highest network SNR and smallest sky localization area.

\begin{table}[t]
\setlength{\tabcolsep}{4pt}
\centering\begin{tabular}{lccccccc} 
\hline
\rule{0pt}{3ex}%
Parameter & GW170104 & GW170608 & GW170729 & GW170809 & GW170814 & GW170818 & GW170823 \\
\hline\hline
\rule{0pt}{3ex}%
\vspace*{0.1cm}
$\mathcal{M}^{\mathrm{det}}$ (M$_{\odot}$) & $25.2^{+1.7}_{-1.6}$ & $8.50^{+0.06}_{-0.05}$ & $51.7^{+8.0}_{-9.0}$ & $29.9^{+2.2}_{-1.8}$ & $27.2^{+1.2}_{-1.2}$ & $32.2^{+2.8}_{-2.8}$ & $39.1^{+4.7}_{-4.5}$ \\
\vspace*{0.1cm}
$m_1^{\mathrm{det}}$ (M$_{\odot}$) & $37.3^{+8.2}_{-6.8}$ & $12.0^{+6.0}_{-2.1}$ & $74.5^{+13.0}_{-13.8}$ & $41.9^{+10.3}_{-6.8}$ & $33.9^{+6.3}_{-2.8}$ & $43.5^{+9.7}_{-6.1}$ & $52.7^{+12.7}_{-8.1}$ \\
\vspace*{0.1cm}
$m_2^{\mathrm{det}}$ (M$_{\odot}$) & $22.9^{+5.9}_{-4.9}$ & $8.0^{+1.6}_{-2.3}$ & $48.8^{+14.6}_{-16.0}$ & $28.7^{+5.9}_{-6.6}$ & $28.9^{+2.6}_{-4.4}$ & $32.0^{+5.9}_{-7.6}$ & $39.1^{+7.8}_{-10.6}$ \\
\vspace*{0.1cm}
$\mathcal{M}^{\mathrm{src}}$ (M$_{\odot}$) & $21.2^{+1.9}_{-1.4}$ & $7.96^{+0.19}_{-0.19}$ & $34.1^{+6.4}_{-4.5}$ & $24.9^{+2.1}_{-1.5}$ & $24.3^{+1.4}_{-1.2}$ & $26.7^{+2.2}_{-1.9}$ & $29.0^{+4.2}_{-3.2}$ \\
\vspace*{0.1cm}
$m_1^{\mathrm{src}}$ (M$_{\odot}$) & $31.4^{+7.6}_{-6.0}$ & $11.3^{+5.6}_{-2.0}$ & $49.5^{+12.1}_{-10.2}$ & $35.0^{+9.1}_{-5.9}$ & $30.4^{+5.7}_{-2.7}$ & $36.1^{+8.5}_{-5.3}$ & $39.2^{+10.9}_{-6.6}$ \\
\vspace*{0.1cm}
$m_2^{\mathrm{src}}$ (M$_{\odot}$) & $19.2^{+4.9}_{-4.0}$ & $7.5^{+1.5}_{-2.2}$ & $32.2^{+9.9}_{-9.1}$ & $23.9^{+5.0}_{-5.3}$ & $25.8^{+2.6}_{-4.0}$ & $26.5^{+4.7}_{-6.0}$ & $28.9^{+6.3}_{-7.2}$ \\
\vspace*{0.1cm}
$q$ & $1.63^{+0.84}_{-0.56}$ & $1.50^{+1.65}_{-0.46}$ & $1.53^{+0.93}_{-0.48}$ & $1.46^{+0.85}_{-0.42}$ & $1.17^{+0.46}_{-0.15}$ & $1.36^{+0.76}_{-0.33}$ & $1.34^{+0.85}_{-0.31}$ \\
\vspace*{0.1cm}
$\chi_{\mathrm{eff}}$ & $-0.08^{+0.16}_{-0.17}$ & $0.057^{+0.19}_{-0.06}$ & $0.34^{+0.21}_{-0.27}$ & $0.06^{+0.18}_{-0.16}$ & $0.08^{+0.12}_{-0.12}$ & $-0.08^{+0.20}_{-0.24}$ & $0.07^{+0.22}_{-0.21}$ \\
\vspace*{0.1cm}
$a_1$ & $0.35^{+0.48}_{-0.31}$ & $0.32^{+0.47}_{-0.29}$ & $0.60^{+0.34}_{-0.51}$ & $0.34^{+0.53}_{-0.31}$ & $0.53^{+0.42}_{-0.48}$ & $0.56^{+0.38}_{-0.50}$ & $0.44^{+0.48}_{-0.40}$ \\
\vspace*{0.1cm}
$a_2$ & $0.47^{+0.45}_{-0.42}$ & $0.43^{+0.49}_{-0.39}$  & $0.57^{+0.38}_{-0.50}$ & $0.40^{+0.51}_{-0.37}$ & $0.46^{+0.47}_{-0.42}$ & $0.50^{+0.44}_{-0.45}$ & $0.45^{+0.48}_{-0.41}$ \\
\vspace*{0.1cm}
$d_L$ (Mpc) & $970^{+400}_{-410}$ & $318^{+128}_{-109}$ & $2980^{+1410}_{-1400}$ & $1020^{+310}_{-390}$ & $584^{+130}_{-186}$ & $1030^{+420}_{-350}$ & $1920^{+870}_{-860}$ \\
\hline
\end{tabular}
\caption{Results from PyCBC Inference analysis of binary black hole events from LIGO-Virgo's second observing run. Quoted are the median and 90\% credible interval values for a subset of the inferred model parameters.}
\label{tab:results}
\end{table}

Estimates of the parameters for these events were previously published in the LIGO--Virgo Collaboration (LVC) detection papers for these events~\cite{Abbott:2017vtc,Abbott:2017gyy,Abbott:2017oio,LIGOScientific:2018mvr}. The results from our analyses are overall in agreement with the estimates published by the LVC within the statistical errors of measurement of the parameters. Any small discrepancies in the measurement of the parameters would be due to the differences in the analysis methods. One of the differences is the method of the PSD estimation. Another such difference is that we do not marginalize over calibration uncertainties of the measured strain~\cite{PhysRevD.96.102001}, whereas the LVC analyses use a spline model to fit the calibration uncertainties. The true impact of calibration errors on the parameter estimates should be evaluated using a physical model of the calibration, which does not exist currently in any analysis. This will be revisited in a future work.

\subsection*{Code availability}
The posterior probability density functions presented in this paper were sampled using the PyCBC Inference software. The PyCBC Inference toolkit uses the Bayesian inference methodology described in this paper; a more detailed description of the toolkit is presented in Ref.~\cite{Biwer:2018osg}. The source code and documentation of PyCBC Inference is available as part of the PyCBC software package at \verb"http://pycbc.org". The results in this paper were generated with the PyCBC version 1.12.3 release. In the data release repository for this work~\cite{data_release} we provide scripts and configuration files for replicating our analysis. The scripts document our command line calls to the \texttt{pycbc\_inference} executable which performs the ensemble MCMC analyses. The command line call to \texttt{pycbc\_inference} contains options for: the ensemble MCMC configuration, data conditioning, and locations of the configuration file and gravitational-wave detector data files. The configuration files included in the repository, and used as an input to \texttt{pycbc\_inference}, specify the prior probability density functions used in the analyses, including sections for: initializing the distribution of Markov-chain positions in the ensemble MCMC, declaring transformations between the parameters that define the prior and the parameters the ensemble MCMC samples (eg. $(m_1, m_2) \rightarrow (\mathcal{M}, q)$), and defining additional constraints to the prior probability density function~\cite{Biwer:2018osg}.

\section*{Data Records}\label{sec:data_record}

The data products from the parameter estimation analyses for the seven events are stored in seven HDF~\cite{andrew_collette_2018_1246321} files, available within the Zenodo data release repository~\cite{data_release} for this work. The location of these HDF files within the repository are listed in Table~\ref{tab:data_rec}. In this section, we describe the contents of these seven HDF files.

The top-level of the HDF files contains attributes named \texttt{ifos}, \texttt{variable\_args}, \texttt{posterior\_only}, and \texttt{lognl}. \texttt{variable\_args} is a list of the inferred model parameters.
For these seven analyses this includes: the coalescence time (\texttt{tc}), distance (\texttt{distance}), inclination angle (\texttt{inclination}), polarization angle (\texttt{polarization}), right ascension (\texttt{ra}), declination (\texttt{dec}), detector-frame component masses (\texttt{mass1} and \texttt{mass2}), azimuthal angles of the spin vector (\texttt{spin1\_azimuthal} and \texttt{spin2\_azimuthal}), polar angles of the spin vector (\texttt{spin1\_polar} and \texttt{spin2\_polar}), and magnitudes of the spin vector (\texttt{spin1\_a} and \texttt{spin2\_a}). \texttt{mass1}, \texttt{spin1\_a}, \texttt{spin1\_polar}, \texttt{spin1\_azimuthal} in the files refer to the primary black hole in the binary. \texttt{mass1}, \texttt{spin2\_a}, \texttt{spin2\_polar}, \texttt{spin2\_azimuthal} refer to the secondary black hole in the binary. 

\texttt{ifos} stores the list of the names of interferometers from which data has been analyzed in each run. The attribute \texttt{posterior\_only} is a Boolean where a \texttt{True} value indicates that the posterior samples and likelihood statistics are stored as flattened arrays in the files. \texttt{lognl} stores the value of the noise likelihood, which is described below.    

The independent samples of the model parameters are stored in a top-level HDF group, named \texttt{[`samples']}. For each parameter listed in the \texttt{variable\_args} attribute, the \texttt{[`samples']} HDF group contains an HDF dataset that is a one-dimensional array indexed by the independent samples. Therefore, the set of parameters for the $i$-th independent sample is the $i$-th element of each array. For example, \texttt{[`samples/mass1'][32]} and \texttt{[`samples/mass2'][32]} are the masses for the 32-nd independent sample. Samples in the \texttt{mass1} and \texttt{mass2} data sets are in solar mass units, those in \texttt{distance} are in Mpc units, those in \texttt{tc} are in seconds, and those in \texttt{spin1\_a} and \texttt{spin2\_a} are dimensionless. Samples in the \texttt{spin1\_polar}, \texttt{spin2\_polar}, \texttt{spin1\_azimuthal}, \texttt{spin2\_azimuthal}, \texttt{inclination}, \texttt{ra}, \texttt{dec}, and \texttt{polarization} are in radians.


The second top-level HDF group is \texttt{[`prior\_samples']}, which stores prior samples in a similar format as the \texttt{[`samples']} group described above. For each of the parameters listed in the \texttt{variable\_args} attribute, the \texttt{[`prior\_samples']} HDF group contains an HDF dataset that is a one-dimensional array of samples of that parameter drawn from the prior distribution.

The third top-level HDF group, named \texttt{[`likelihood\_stats']}, contains quantities to obtain the prior $\prior$ and likelihood $\likelihood$ from Eq.~\ref{eqn:bayes} for each independent sample.
In order to obtain the prior for each independent sample, the \texttt{[`likelihood\_stats']} HDF group contains a dataset of the natural logarithm of the prior probabilities called \texttt{[`likelihood\_stats/prior']}. The datasets in the \texttt{[`likelihood\_stats']} HDF group are one-dimensional arrays indexed by the independent sample (eg. the $i$-th element corresponds to the prior probability of the $i$-th independent sample) as well. In order to obtain the likelihood for each independent sample, there is a dataset containing the natural logarithm of the likelihood ratio $\Lambda$ called \texttt{[`likelihood\_stats/loglr']}. The likelihood ratio $\Lambda$ is defined as~\cite{Biwer:2018osg}
\begin{equation}\label{eqn:lambda}
\log \Lambda = \log \frac{\likelihood}{p(\vec{d}(t)|\vec{n})}
\end{equation}
where $\log p(\vec{d}(t)|\vec{n})$ is the natural logarithm of the noise likelihood defined as~\cite{Biwer:2018osg}
\begin{equation}\label{eqn:noise_likelihood}
\log p(\vec{d}(t)|\vec{n}) = - \frac{1}{2} \sum_{i=1}^{N} \left< \tilde{d}_{i}(f) | \tilde{d}_{i}(f) \right> \,.
\end{equation}
The natural logarithm of the noise likelihood is a constant for each analysis. Therefore from Eq.~\ref{eqn:lambda}, in order to compute the natural logarithm of the likelihood, $\log \likelihood$, the user adds \texttt{lognl} to each element of \texttt{[`likelihood\_stats/loglr']}.

The fourth top-level HDF group is \texttt{[`psds']}. For each interferometer from which data has been used in the analysis, the \texttt{[`psds']} HDF group contains a dataset storing a frequency series of the PSD multiplied by the square of the dynamic range factor. The dynamic range factor is a large constant to reduce the dynamic range of the strain; here, we use $2^{69}$ rounded to 17 significant figures (precisely $5.9029581035870565\times 10^{20}$). The first entry in each PSD frequency series corresponds to frequency $f = 0$~Hz, and the last entry corresponds to $f = 1024$~Hz. Attached as attributes to each interferometer's PSD frequency series dataset object are the frequency resolution---\texttt{delta\_f} and the low frequency cutoff used for that interferometer in the PSD estimation and likelihood computation---\texttt{low\_frequency\_cutoff}.

\vspace{0.5cm}
\begin{table*}[t]
\begin{tabular}{ |p{2.5cm}||p{1.5cm}|p{10.5cm}|  }
 \hline
 Event & Posterior samples & Location of the associated data file within the data release repository \\
 \hline
 \hline
 GW170104 & 8000 & \verb"posteriors/GW170104/gw170104_posteriors_thinned.hdf" \\
 \hline
 GW170608 & 8000 & \verb"posteriors/GW170608/gw170608_posteriors_thinned.hdf" \\
 \hline 
 GW170814 & 8000 & \verb"posteriors/GW170814/gw170814_posteriors_thinned.hdf" \\
 \hline
 GW170729 & 8000 & \verb"posteriors/GW170729/gw170729_posteriors_thinned.hdf" \\
 \hline
 GW170809 & 8000 & \verb"posteriors/GW170809/gw170809_posteriors_thinned.hdf" \\
 \hline
 GW170818 & 8000 & \verb"posteriors/GW170818/gw170818_posteriors_thinned.hdf" \\
 \hline
 GW170823 & 8000 & \verb"posteriors/GW170823/gw170823_posteriors_thinned.hdf" \\
 \hline
\end{tabular}
\caption{For each binary black hole merger, this table contains: the event's name, number of independent samples obtained with the ensemble MCMC, and location of the HDF files containing the independent samples within the Zenodo data release repository~\cite{data_release}.\label{tab:data_rec}}
\end{table*}
\vspace{0.5cm}

\section*{Technical Validation}\label{sec:validation}

The analyses in this paper were performed using the PyCBC Inference software~\cite{Biwer:2018osg} with the parallel-tempered emcee sampler~\cite{emcee,mcmc} (\verb"https://github.com/dfm/emcee/tree/v2.2.1"), hereafter referred to as \texttt{emcee\_pt}, as the sampling algorithm. A validation study of PyCBC Inference with the \texttt{emcee\_pt} sampler was presented in Sec.~4 of Ref.~\cite{Biwer:2018osg}. The validation study in Ref.~\cite{Biwer:2018osg} used the same version of the PyCBC code, waveform model, sampler settings, data conditioning settings, and burn-in test as used in our analyses in this paper, and therefore demonstrates the credibility of the results presented in this paper. In this section, we summarize the validation study.

We have tested the performance of this setup (ie. code version, waveform model, sampler settings, etc.) using analytic likelihood functions such as the multivariate normal, Rosenbrock, eggbox, and volcano functions. The \texttt{emcee\_pt} sampler successfully sampled the underlying analytical distributions. The recovery of parameters of a four-dimensional normal distribution using the \texttt{emcee\_pt} sampler is shown in Fig.~2 of Ref.~\cite{Biwer:2018osg}.

Ref.~\cite{Biwer:2018osg} also describes a test performed using simulated binary black hole signals to validate the reliability of parameter estimates generated by PyCBC Inference with the \texttt{emcee\_pt} sampler. The test is carried out by generating 100 realizations of stationary Gaussian noise colored by the power spectral densities of the Advanced LIGO detectors around the time of observation of GW150914~\cite{Abbott:2016blz}. A unique simulated binary black hole signal, whose parameters were sampled from the prior probability density function, is injected into each simulated noise realization. For the population of 100 simulated binary black hole signals, the network signal-to-noise ratios range from 5 to 160, and are predominantly spaced between 10 to 40. PyCBC Inference, using the \texttt{emcee\_pt} sampler, was then run on each simulated binary black hole signal to produce samples of the posterior probability density function and compute credible intervals that estimate the modeled parameter values. For each parameter, we then calculate the percentage of the runs ($x\%$) in which the true value of the parameter was recovered within a certain credible interval ($y\%$). In the ideal case, there should be a 1-to-1 relation between these percentiles, ie. $x$ should equal $y$ for any value of the percentile $y$. The percentile-percentile curves obtained for each parameter in the test is plotted in Fig.~3 of Ref.~\cite{Biwer:2018osg}. To evaluate the deviation between the percentile-percentile curve for each parameter from a 1-to-1 relation, a Kolmogorov-Smirnov (KS) test is performed. Using the set of p-values obtained for all the parameters, another KS test is performed expecting the p-values to adhere to a uniform distribution. The p-value obtained from this calculation is 0.7, which is sufficiently high to infer that PyCBC Inference, with it's implementation of the \texttt{emcee\_pt} sampler, provides unbiased estimates of the binary black hole modeled parameters.

In addition to the aforementioned tests using analytical distributions and simulated signals, the 90\% credible interval measurements of the binary black hole parameters from our analyses presented in this paper are in agreement with the LIGO--Virgo Collaboration estimates~\cite{Abbott:2017vtc,Abbott:2017gyy,Abbott:2017oio} which used a different inference code. This further validates the results presented here.

\section*{Usage Notes}\label{sec:con}
When citing the data associated with this paper and released in the data release repository~\cite{data_release}, please cite this paper for describing the data and the analyses that generated them. Please also cite Ref.~\cite{Biwer:2018osg} which describes and validates the PyCBC Inference parameter estimation toolkit that was used for generating the data. The samples of the posterior probability density function for each analysis presented in this paper are stored in separate HDF files, and the location of each HDF file is listed in Table~\ref{tab:data_rec}. We direct users to the tools available in PyCBC Inference to read these files and visualize the data. Figs.~\ref{fig:m1m2_plots}, \ref{fig:qchieff_plots}, and \ref{fig:dliota_plots} in this paper were generated using these tools from the PyCBC version 1.12.3 release. The data release repository also includes scripts to execute \texttt{pycbc\_inference} and reproduce the analysis and resulting samples.

The data release repository for this work~\cite{data_release} includes two IPython notebooks named \newline \verb"data_release_o2_bbh_pe.ipynb" and \verb"o2_bbh_pe_skymaps.ipynb". \verb"data_release_o2_bbh_pe.ipynb" \newline presents tutorials for using PyCBC to handle the data. This notebook contains examples to load the HDF datasets, convert the parameters in the HDF files to other coordinates (eg. $(m_1^{\mathrm{det}}, m_2^{\mathrm{det}}) \rightarrow (\mathcal{M}^{\mathrm{det}}, q)$), and visualize the samples of the posterior probability density function. The samples' credible intervals are visualized as marginalized one-dimensional histograms and two-dimensional credible contour regions. We include commands in this notebook to reproduce Figs.~\ref{fig:m1m2_plots}, \ref{fig:qchieff_plots}, and~\ref{fig:dliota_plots} in this paper.
PyCBC Inference also includes an executable called \texttt{pycbc\_inference\_plot\_posterior} to render these visualizations. 
The IPython notebook \verb"o2_bbh_pe_skymaps.ipynb" demonstrates a method of visualizing the sky location posterior distributions, as presented in Fig.~\ref{fig:skymap_plots} in this paper. We use tools from the open source ligo.skymap package (\verb"https://pypi.org/project/ligo.skymap/") for writing the sky location posterior samples from our analyses into FITS files, reading them, and generating probability density contours on a Mollweide projection.

The released data are freely available under the Creative Commons License: CC BY.

\section*{Acknowledgements} \label{sec:ack}
This research has made use of data obtained from the Gravitational Wave Open Science Center \newline (\verb"https://www.gw-openscience.org"), a service of LIGO Laboratory, the LIGO Scientific Collaboration and the Virgo Collaboration. LIGO is funded by the U.S. National Science Foundation. Virgo is funded by the French Centre National de Recherche Scientifique (CNRS), the Italian Istituto Nazionale della Fisica Nucleare (INFN) and the Dutch Nikhef, with contributions by Polish and Hungarian institutes. Computations were performed in the Syracuse University SUGWG cluster.

\textbf{Funding:} This work was supported by NSF awards PHY-1707954 (DAB, SD), and PHY-1607169 (SD). SD was also supported by the Inaugural Kathy '73 and Stan '72 Walters Endowed Fund for Science Research Graduate Fellowship at Syracuse University. Computations were supported by Syracuse University and NSF award OAC-1541396.

\section*{Author contributions} Conceptualization: DAB; Methodology: SD, CMB, CDC, AHN; Software: CMB, CDC, SD, AHN, DAB; Validation: CDC, CMB, AHN; Formal Analysis: SD; Investigation: SD, CMB, CDC, AHN; Resources: DAB; Data Curation: DAB, CDC, CMB, AHN, SD; Writing: SD, CMB, CDC, DAB, AHN; Visualization: SD, CMB, CDC, AHN; Supervision: DAB; Project Administration: DAB; Funding Acquisition: DAB.

\section*{Competing interests}
The authors declare no competing financial interests.




\end{document}